\documentstyle [12pt]{article}
\newcommand{\be}{\begin{equation}}
\newcommand{\ee}{\end{equation}}
\newcommand{\bea}{\begin{eqnarray}}
\newcommand{\ena}{\end{eqnarray}}
\newcommand{\beas}{\begin{eqnarray*}}
\newcommand{\enas}{\end{eqnarray*}}

\newcommand{\g}{U_qg}

\newcommand{\D}{\Delta}
\newcommand{\ot}{\otimes}

\newcommand{\nn}{\nonumber}
\newcommand{\la}{\lambda}

\newcommand{\hot}{\hat{\otimes}}

\newcommand{\op}{\oplus}
\renewcommand{\H}{\bar{H}}
\newcommand{\R}{\hat{R}}
\newcommand{\ab}[1]{{\rm{\bf #1}}}
\renewcommand{\dim}{{\rm dim}}
\newcommand{\id}{{\rm id}}

\begin{document}


\rightline{BUTP-95/20}

\begin{center}

{\large {\bf Spin Chain Hamiltonians with Affine $U_q g$ symmetry}}

\vspace{24pt}

            T.Hakobyan
      \footnote {e-mail:{\sl hakob@vx1.yerphi.am}}

\vspace{10pt}

{\sl Yerevan Physics Institute,}\\
{\sl Br.Alikhanian st.2,375036, Yerevan,Armenia}

\vspace{10pt}

            A.Sedrakyan
           \footnote{Permanent address: {\sl Yerevan Physics Institute,
            375036, Yerevan, Armenia}

           {\sl e-mail: sedrak@vx1.yerphi.am}}

\vspace{10pt}

{\sl Institute for Theoretical Physics,}\\
{\sl Sidlerstasse 5, CH-3012, Bern, Switzerland}

\vspace{15pt}

            June 1995

\end{center}

\vspace{2cm}

\begin{abstract}
We construct the family of spin chain Hamiltonians, which have
 affine $U_q g$ guantum group symmetry. Their eigenvalues
coincides with
the eigenvalues of the usual spin chain Hamiltonians which have
non-affine $U_q g_0$ quantum group symmetry, but have the degeneracy
of levels, corresponding to affine $U_q g$. The space of states
of these chaines are formed by the tensor product of the fully
reducible representations.
\end{abstract}

\newpage

\section{Introduction}

Quantum group symmetry plays great role in integrable statistical
models \cite{J85,J86_1,Dr86} and conformal field theory \cite{MR,PS,GS91}.

It is well known that many integrable Hamiltonians have quantum
group symmetry.
For example, $XXZ$ Heisenberg Hamiltonian with particular boundary terms
\cite{PS}
is $U_qsl_2$-invariant. Infinite $XXZ$ spin chain has larger symmetry:
affine $U_q\widehat{sl}_2$ \cite{DFJMN93}. Single spin site of
most considered Hamiltonians form irreducible representation of
Lie algebra or its quantum deformation.

Here we conctruct family of spin chain Hamiltonians, which have
affine quantum group symmetry. The space of states of these chains are
formed by the tensor product of the fully reducible representations.
We show that the
model, considered in \cite{RA}, which corresponds to some generalization
 of the Habbard Hamiltonian in the strong repulsion
limit, is a particular case of our general construction.
The affine quantum group symmetry leads to high degeneracy of energy
levels.

The energy levels of these spin chains are formed on the states,
constructed from highest weight vectors of quantum group representations.
In particular cases the restriction of considered spin chain
on these states gives rise
to Heisenberg spin chain or Haldane Shastry long range interaction
spin chain.

It is difficult in a moment to name a set of physical problems,
with which the constructed Hamiltonians dirrectly related(besides
mensioned). However it is essential to point out that affine
symmetries appear
in 2D physics when matter fields interact with gravity( in a
noncritical string theory).

\section{Definitions}

Let us recall the definition of quantum Kac-Moody group $\g$. It is
generated by the generators $e_i, \ f_i, \ h_i$ satisfying the relations
\beas
&{[}h_i,e_j{]}  = c_{ij}e_j \qquad {[}h_i,f_j{]}  = -c_{ij}f_j&
 \label{eq:defhg} \\
& {[}e_i,f_j{]} = \delta_{ij}{[}h{]}_q &
\enas
and $q$-deformed Serre relations, which we don't write here.
Here $q$ is a deformation parameter, ${[}x{]}_q:=
(x^q-x^{-q})/(q-q^{-1})$, $c_{ij}$ is a Cartan matrix of
corresponding Kac-Moody algebra $g$.

On $\g$ there is a Hopf algebra structure:
\beas
\D(e_i) =  k_i \ot e_i+e_i\ot k_i^{-1} &
\D(k_i^{\pm1}) = k_i^{\pm1} \ot k_i^{\pm1}
\nn\\
\D(f_i) = k_i \ot f_i+f_i\ot k_i^{-1}
\label{eq:defcomul}
\enas
where $k_i:=q^{h_i\over 2}$.
This comultiplication can be extent to $L$-fold tensor
product by
\beas
\D^{L-1}(e_i) = \sum_{l=1}^L
	k_i \ot\ldots\ot k_i\ot\underbrace{ e_i}_l\ot k_i^{-1}
	\ot\ldots\ot k_i^{-1} \nn\\
\D^{L-1}(f_i) = \sum_{l=1}^L
	k_i \ot\ldots\ot k_i\ot\underbrace{ f_i}_l\ot k_i^{-1}
	\ot\ldots\ot k_i^{-1} \nn\\
\D^{L-1}(k_i^{\pm1}) = k_i^{\pm1} \ot\ldots\ot k_i^{\pm1}\nn
\label{eq:Ldefcomul}
\enas

Let $g$ be an affine algebra and $g_0$ is underlying finite
algebra: $g=\hat g_0$. Then
for any complex $x$ there is the $q$-deformation of
loop homomorphism $\rho_x$:
$\g\rightarrow\g_0$, which is given by
\be
\begin{array}{lll}
\rho_x (e_0) = xf_\theta & \rho_x (f_0)= x^{-1}e_\theta &
	 \rho_x(h_0)=-h_\theta \\
\rho_x (e_i) = e_i & \rho_x (f_i) = f_i & \rho_x (h_i) = h_i,
\end{array}
\nn
\ee
where $i=1\dots n$ and
$\theta$ is a maximal root of $\g$. Using  $\rho_x$ one can construct
the spectral parameter dependent representation  of $\g$  from
the   representation   of $\g_0$.

Let $V_1(x_1)$ and $V_2(x_2)$ are constructed in such way irreducible finite
dimensional
representations of $\g$ with parameters $x_1$ and $x_2$ correspondingly.
 The $\g$-representations on $V_1(x_1)\ot V_2(x_2)$ constructed
by means of $\D$ and $\bar{\D}$ are both irreducible, in general, and
equivalent:
\be
R(x_1,x_2)\D(g)=\bar{\D}(g)R(x_1,x_2), \quad g\in \g
\label{Rmatrix}
\nn
\ee
The $R$-matrix $R(x_1,x_2)$ depends only on $x_1/ x_2$ and is a Boltsmann
weight of some integrable statistic mechanical system.

\section{Quantum group invariant Hamiltonians for reducible
        representations}

Let $V=\oplus_{i=1}^N V_{\lambda_i}$ is a direct sum of finite dimensional
irreducible representations of $\g$.
We denote by $V(x_1,\dots,x_N)$ corresponding affine $\g$
representation with spectral parameters $x_i$:
$$
V(x_1,\dots,x_N)=\oplus_{i=1}^NV_{\lambda_i}(x_i)
$$
We consider the intertwining operator
$$
H(x_1,\dots,x_N): \ V(x_1,\dots,x_N)\ot V(x_1,\dots,x_N)\rightarrow
V(x_1,\dots,x_N)\ot V(x_1,\dots,x_N),
$$
${[}H(x_1,\dots,x_N),\D(a){]}=0$, for all $a\in\g$.
If $V=V_\lambda$
consists of one irreducible component then $H$ is a multiple of identity,
because the tensor product is irreducible in this case.
To carry out the general case let
us gather all equivalent irreps together:
\footnote{The $\g$-equivalence of $V_{\lambda_i}(x_i)$ requires
that the spectral parameters
$x_i$ and highest weights $\lambda_i$ are the same. }
$$
 V(x_1,\dots,x_N)=\bigoplus_iN_{\la_i}\hot V_{\la_i}(x_i),
$$
where all $V_{\la_i}(x_i)$ are nonequivalent and
$N_{\la_i}\simeq \ab{C}^{n_i}$ have a dimension equal to
the multiplicity of $V_{\la_i}(x_i)$
in $V(x_1,\dots,x_N)$.
By the hat over the tensor product we mean that $\g$ doesn't act on
$N_{\la_i}\hot V_{\la_i}(x_i)$ by means of $\D$ but acts as $id\ot g$.

So, we have:

\bea
\label{decompos}
 V(x_1,\dots,x_N)\ot V(x_1,\dots,x_N)
	&=&(\oplus_iN_{\la_i}\hot V_{\la_i}(x_i))\bigotimes
	(\oplus_iN_{\la_i}\hot V_{\la_i}(x_i))  \nonumber\\
	&=&
	\bigoplus_{i,j}N_{\la_i}\hot N_{\la_j}\hot
	\left( V_{\la_i}(x_i)\ot V_{\la_j}(x_j)\right)
\ena

Now, $ V_{\la_i}(x_i)\ot V_{\la_j}(x_j)$ is equivalent only to itself and to
$ V_{\la_j}(x_j)\ot V_{\la_i}(x_i)$ (for $i\ne j$) by the operator
$\R(x_i/x_j)=PR(x_i/x_j)$,
where $P$ is tensor product permutation: $P(v_1\ot v_2)=v_2\ot v_1$.
So,
the commutant $H(x_1,\dots,x_N)$
of $\g$ on $V(x_1,\dots,x_N)\ot V(x_1,\dots,x_N)$
has the following form:

\be\label{Hact}
H|_{\bigoplus_{i,j}N_{\la_i}\hot N_{\la_j}\hot V_{\la_i}\ot V_{\la_j}}=
	A_{ij}\hot id_{V_{\la_i}\ot V_{\la_j} } +
	B_{ij}\hot \R_{V_{\la_i}\ot V_{\la_j} }(x_i/x_j)
\ee
where $A_{ij}$ and $B_{ij}$ are any operators on $N_{\la_i}\hot N_{\la_j}$

Let us consider some particular cases of this general construction.
\begin{enumerate}
 \item Let $V(x)=V(x,x)=V_\la(x)\oplus V_\la(x)$.
	The second term in (\ref{Hact}) is absent in this case and $H$ has
	factorized form:
\bea
H=A\hot id_{V_\la\ot V_\la},
     &  A=a^{\alpha\delta}_{\beta\gamma}
\nn
\ena
	where $\alpha,\beta,\gamma,\delta=\pm$ are indexes, corresponding to
	each
	$V_\la$.
\item
Let now $V(x_1,x_2)=V_{\la_1}(x_1)\oplus V_{\la_2}(x_2)$ ($V_{\la_i}(x_i)$
 are mutually
nonequivalent). Then $H$ acquires the following form
\be
\label{R}
H(x_1,x_2)= \left(
	\begin{array}{cccc}
		 a\cdot\id & 0 & 0 & 0 \\
     	0 & c\cdot\id & d\cdot R_{21}(x_2/x_1)& 0\\
 		0 & e\cdot R_{12}(x_1/x_2) & f\cdot\id & 0\\
     	 0 & 0 & 0 & g\cdot\id
     \end{array}
\right)
\nn
\ee
Here we used $R_{21}=\sum_i b_i\ot a_i$ for $R_{12}=\sum_i a_i\ot b_i$.
Note, that we can normalize $R$-matrices to satisfy the unitarity
condition
$
R_{12}(z)R_{21}(z^{-1})=\id
$.
This leads to
\be
\label{inv}
H(x_1,x_2)^2=\id\ot\id
\ee

\item
If we choose $g=sl(2)$ and $V=V_{1\over 2}\op V_0\op V_0 \op \ldots
\op V_0$, where $V_{1\over 2}$ is fundamental representation of
$U_qsl_2$ and $V_0$ is trivial one dimensional representation of one,
one can obtain the Hamiltonian, corresponding to a strong repulsion limit
of some generalization of Habbard model, considered in \cite{RA}.
The representation (5.13) there is a $U_qsl_2$-representation on $V$.
\end{enumerate}

Following \cite{RA} from the operator $H$ the following Hamiltonian
acting on $W=V^{\ot L}$ can be constructed:
\footnote{Here and in the following we omit the dependence on $x_i$}
\be
\H=\sum_{i=1}^{L-1}H_{ii+1}
\label{genhamil}
\ee
Here and in the following for the operator $X=\sum_l x_l\ot y_l$ on
$V\ot V$ we denote by $X_{ij}$ its action on $W$ defined by
\be
\label{ij}
 X_{ij}=\sum_l \id\ot\dots \ot \id\ot
\underbrace{x_l}_{i}\ot \id\dots\ot\id\ot \underbrace{y_l}_{j}\ot\id\ot
	\dots\ot \id
\ee
By the construction, $\H$ is quantum group invariant:
\bea
{[}\H,\D^{L-1}(g){]}=0 & \forall g\in \g
\nn
\ena

Let $V^0$ is the linear space, spanned by the highest
weight vectors in $V$: $V^0:=\op_{i=1}^N v_{\la_i}^0$, where $v_{\la_i}\in
V_{\la_i}$ is a highest weight vector, and $W^0:=V^{0\ \ot L}$. The space
$W^0$ is $\H$-invariant. This
follows from the intertwining property of $\H$.
For general $q$, $W$ is $\g$-irreducible module
so the action of $\g$ on $W^0$ generate all $W$. So, the energy levels
of $\H$ are highly degenerate.

First, one can consider $\H$ on the space
$W^0$ and determine (if it is possible)  the energy levels and corresponding
eigenvectors there. Then performing the quantum group on each
eigenvector of some energy level one can obtain the whole eigenspace for
this level. Moreover, the space $W^0$ itself is a direct sum of
$\H$-invariant spaces, each is spanned by the tensor products of fixed number
highest weight vectors from each equivalence class of irreps:

\beas
\label{w0decompos}
&W^0=\bigoplus_{\stackrel{p_1,\dots,p_M}{p_1+\dots+p_M=L}}W^0_{p_1\dots p_M}&
\nn\\
&W^0_{p_1\dots p_M}:=
     \left\{ \bigoplus \ab{C} v_{\la_{i_1}}^0\ot\ldots\ot v_{\la_{i_L}}^0
 |  \# (\la_i,x_i)\in
        \{(\la_1,x_1),\dots,(\la_N,x_N)\}=p_i\right\}&
\enas

The $\H$-invariance of $W^0_{p_1\dots p_M}$ follows again from the
definition of $\H$ as an intertwining operator. The energy levels are
now determined on these spaces.  Note that the dimension of
$W^0_{p_1\dots p_M}$  is
$$\left(\begin{array}{c} L \\ p_1\ldots p_M \end{array}\right)$$

Every Hamiltonian eigenvector
$w_0\in
W^0_{p_1\dots p_M}
$
gives rise to a $\g$-representation space of dimension
\be
\label{lev}
\prod_{k=1}^M(\dim V_{\la_k})^{p_k}
\ee
 This is the degeneracy level
of its energy value.
In the
particular case when all $V_{\la_i}$ are equivalent, the degeneracy level
is  $(\dim V_\la)^L$. Note that
\bea
\dim W =\sum_{\stackrel{p_1\dots,p_M}{p_1+\dots+ p_M=L}}
	\left(\begin{array}{c} L \\ p_1\ldots p_N \end{array}\right)
        \prod_{k=1}^N (\dim V_{\la_k})^{p_k} =
        \left(\sum_{k=1}^N N_{\la_k}\dim V_{\la_k}\right)^L
\nn
\ena
as it must be.

For example, if we choose two equivalent representations (the first
case  above),
then $\dim V^0=2$ and there is one term in decomposition (\ref{w0decompos}).
$H$ now is the most general action on $V^0\ot V^0$. As a particular case,
the $XYZ$ Hamiltonian in the magnetic field can be obtained. This
case is most trivial because the degeneracy of all energy levels is the
same.
So, for the statistical sum $Z_{\H}(\beta)=\sum_n\exp(-\beta E_n)$ we have
$$
Z_{\H}(\beta)=(\dim V_\lambda)^L Z_{XYZ}(\beta)
$$

Let us choose
\bea
\label{ex2}
& a=g=e=d=1 \qquad c=f=0 &
\ena
for the  second example. Then the
restriction of $\H$ on $W^0$ coincides with the Bethe $XXX$ spin
chain
\be
\label{XXX}
\H|_{W^0}=H_{XXX}=\sum_i P_{ii+1}={1\over 2}
	\sum_i \left({1+\vec{\sigma}_i \vec{\sigma}_{i+1}}\right)
\ee
The space $W^0_{p_1p_2}$, $p_1+p_2=L$ corresponds to all states
with the same $s_z=p_1/2$ value of spin projection
$
S^z={1/ 2}\sum_i \sigma^z_i
$.
If we return to $\H$ the  energy level degeneracy of each eigenstate with
the same
spin projection is multiplies by
$(\dim V_{\lambda_1})^{2s_z}
(\dim V_{\lambda_2})^{L-2s_z}$.

\section{Generalization to long range interaction spin chains}

Let us consider now the generalization of above construction in case
of long range interacting Hamiltonians.

Recall the Haldane-Shastry
spin chain is given by \cite{Haldane88,Shastry88,Inoz90}
\be
\label{hal}
H_{HS}=\sum_{i<j} {1 \over d_{i-j}^2} P_{ij},
\ee
Here the spins  take values in the fundamental representation
of $sl_n$.
It is well known that the Hamiltonian (\ref{hal}) is integrable if
$d_i$  has one of the following  values
\be
d_j=\left\{
        \begin{array}{ll}
         j, & {\rm rational\ case}\\
        (1/\alpha)\sinh(\alpha j),\ \alpha\in {\bf{\rm R}}, &
		{\rm hyperbolic\ case}   \\
        (L/\pi)\sin(\pi j/L), & {\rm trigonometric\ case}
        \end{array}
\right.
\ee
The trigonometric model is defined on periodic chain and the sum in
(\ref{hal})  is performed over
$1\le i,j\le L$.   Rational and
hyperbolic models are defined on infinite chain.

One can try to generalize the Hamiltonian (\ref{hal}) for
the reducible spin representations by
\be
\H_{HS}=\sum_{i<j} {1 \over d_{i-j}^2} H_{ij},
\ee
where $H$ is taken for the case (\ref{ex2}) of second
example in the previous section.
But it is easy to see that it isn't invariant with respect to
quantum group. This is because the equation
\be
\R_{ij}(x_1,x_2)\D^{L-1}(g)=\D^{L-1}(g)\R_{ij}(x_1,x_2), \quad g\in \g
\ee
is valid only for $i=j\pm1$.

To overcome this difficulty let us substitute instead of $H_{ij}$
the operator
\footnote{
Note that $F_{[ij]}$ and $G_{[ij]}$ act nontrivially on all indexes
$i,i+1,\dots,j$. So we include them into bracket to not
confuse with the definition (\ref{ij}).}
\bea
\label{F}
& F_{[ij]}=G_{[ij]}H_{j-1j}G_{[ij]}^{-1},
\quad {\rm where} \quad
G_{[ij]}=H_{ii+1}
H_{i+1i+2}\ldots H_{j-2j-1} &
\ena
The 'nonlocal' term like $F_{[ij]}$ appeared as a boundary
term in the construction of quantum group invariant and in some
sense periodic spin chains \cite{Martin,GPPR}.

Note that it follows from (\ref{R},\ref{ex2},\ref{inv}) that $H_{ii+1}$
satisfy
\bea
& H_{i-1i}H_{ii+1}H_{i-1i}= H_{ii+1}H_{i-1i}H_{ii+1}
\qquad H_{ii+1}^2=1 & \nn
\ena
This is a realization  of permutation algebra. In contrast to
standard realization by $P_{ij}$,  the relation
\[
P_{i-1i}P_{ii+1}P_{i-1i}=P_{i-1i+1}
\]
isn't fulfilled. The restriction of $H_{ii+1}$ on the highest
weight space $W_0$ coincides with $P_{ii+1}$. Also it is easy to see
from (\ref{F}) that
\[
F_{[ij]}|_{W_0}=P_{ij}
\]
So, the spin chain defined by
\be
\label{genhal}
\H_{HS}=\sum_{i<j} {1 \over d_{i-j}^2} F_{[ij]},
\ee
is quantum group invariant and its restriction on the space
$W^0$ it coincides with the Haldane-Shastry spin chain (\ref{hal}).
The energ
y levels of $\H_{HS}$ coincides with the levels of
(\ref{hal}). The degeneracy degree with respect to the later
is defined by (\ref{lev}).

\section{Aknowlegement}

One of us (A.S.) aknowledge the Institute of Theoretical Physics
in Bern for hospitality and especially H.Leutwyler for many interesting
discussions.

This work was supported by Schweizerischer Nationalfonds
and  Grant 211-5291 YPI of the German Bundesministerium fur
Forschumg und Technologie

\end{document}